%% file: main.tex
\pdfoutput=1

\documentclass[acmsmall,screen,nonacm]{acmart}

\usepackage{multirow}
\usepackage{listings}
\usepackage{subcaption}
\usepackage{wrapfig}
\usepackage{placeins}   %
\usepackage{algorithm}
\usepackage{colortbl}
\usepackage{array}
\usepackage{tabularx}
\usepackage{needspace}
\usepackage{algorithmic}

\definecolor{jacgreen}{RGB}{204,255,204}
\definecolor{jacyellow}{RGB}{255,255,204}
\definecolor{jacred}{RGB}{255,204,204}

\definecolor{magicFull}{HTML}{CCFFCC}
\definecolor{magicPartial}{HTML}{FFFFCC}
\definecolor{magicAbsent}{HTML}{FFCCCC}
\newcommand{\magicYes}{\cellcolor{magicFull}\ensuremath{\checkmark}}
\newcommand{\magicSome}{\cellcolor{magicPartial}\ensuremath{\sim}}
\newcommand{\magicNo}{\cellcolor{magicAbsent}\ensuremath{\times}}

\lstdefinelanguage{Jac}{
  keywords={node,obj,has,def,by,llm,sem,can,with,entry,visit,else,if,not,here,self,str,walker,edge,root,spawn},
  keywordstyle=\color{blue!70!black}\bfseries,
  comment=[l]{\#},
  commentstyle=\color{gray}\itshape,
  string=[b]",
  stringstyle=\color{red!60!black},
  sensitive=true,
  morestring=[b]',
}
\lstdefinestyle{python}{
  language=Python,
  keywordstyle=\color{blue!70!black}\bfseries,
  commentstyle=\color{gray}\itshape,
  stringstyle=\color{red!60!black},
}
\lstdefinestyle{jac}{language=Jac}

\lstdefinestyle{agentcode}{style=python}
\input{macros.tex}

\input{support/listings}

\begin{document}

\title[Jac for Agentic AI]{Agent-Native by Design: A Programming Language for Agentic Applications}

\author{Jayanaka L. Dantanarayana}
\orcid{0009-0000-4320-8280}
\affiliation{%
  \institution{University of Michigan}
  \city{Ann Arbor}
  \state{MI}
  \country{USA}}
\email{jayanaka@umich.edu}
\author{Savini Kashmira}
\orcid{0009-0005-4911-7597}
\affiliation{%
  \institution{University of Michigan}
  \city{Ann Arbor}
  \state{MI}
  \country{USA}}
\email{savinik@umich.edu}
\author{Xiaoyu Ma}
\orcid{0009-0003-5605-4729}
\affiliation{%
  \institution{Shanghai Jiao Tong University}
  \city{Shanghai}
  \country{China}}
\email{xiaoyu\_ma@sjtu.edu.cn}
\author{Malitha Prabhashana}
\affiliation{%
  \institution{Jaseci Labs}
  \country{USA}}
\email{malitha@jaseci.org}
\author{Lingjia Tang}
\orcid{0000-0002-5609-7775}
\affiliation{%
  \institution{University of Michigan}
  \city{Ann Arbor}
  \state{MI}
  \country{USA}}
\email{lingjia@umich.edu}
\author{Jason Mars}
\orcid{0000-0002-7029-5292}
\affiliation{%
  \institution{University of Michigan}
  \city{Ann Arbor}
  \state{MI}
  \country{USA}}
\email{profmars@umich.edu}

\renewcommand{\shortauthors}{Dantanarayana et al.}

\input{sections/abstract}


\maketitle

\captionsetup[figure]{name=Figure}   %

\input{sections/intro}
\input{sections/background}
\input{sections/design}
\input{sections/compiler_runtime}
\input{sections/eval}

\input{sections/related}
\input{sections/conclusion}

\FloatBarrier

\input{sections/statements}

\bibliographystyle{ACM-Reference-Format}
\bibliography{references}

\end{document}

%% file: macros.tex
\newcommand{\sysname}{Jac}

\newif\ifdraftnotes
\draftnotesfalse

\ifdraftnotes
  \newcommand{\todo}[1]{\textcolor{red}{[TODO: #1]}}
  \newcommand{\note}[2]{\textcolor{blue}{[#1: #2]}}
\else
  \newcommand{\todo}[1]{}
  \newcommand{\note}[2]{}
\fi

%% file: support/listings.tex
\usepackage{listings}
\usepackage{xcolor}

\definecolor{codebg}{RGB}{255,253,240}
\definecolor{codeframe}{RGB}{200,200,200}
\definecolor{keyword}{RGB}{0,0,180}
\definecolor{string}{RGB}{163,21,21}
\definecolor{comment}{RGB}{0,128,0}
\definecolor{identifier}{RGB}{0,0,0}
\definecolor{emphcolor}{RGB}{255,140,0} %

\definecolor{outbg}{RGB}{245,248,255}
\definecolor{outframe}{RGB}{200,200,200}
\definecolor{outtext}{RGB}{40,40,40}

\lstdefinestyle{mypython}{
    language=Python,
    backgroundcolor=\color{codebg},
    basicstyle=\ttfamily\scriptsize,
    keywordstyle=\color{keyword}\bfseries,
    stringstyle=\color{string},
    commentstyle=\itshape\color{comment},
    identifierstyle=\color{identifier},
    numbers=left,
    numberstyle=\tiny\color{gray},
    stepnumber=1,
    numbersep=8pt,
    frame=single,
    rulecolor=\color{codeframe},
    tabsize=4,
    showstringspaces=false,
    showspaces=false,
    showtabs=false,
    breaklines=true,
    breakatwhitespace=false,
    keepspaces=true,
    columns=fullflexible,
    linewidth=\linewidth,
    xleftmargin=0.5cm,
    xrightmargin=0.1cm,
    breakindent=0pt,
    captionpos=b,
    emph={Tensor,torch,nn,autograd},
    emphstyle=\color{emphcolor}\bfseries,
    postbreak={}
}

\lstnewenvironment{python}[1][]{
    \lstset{style=mypython,#1}
}{}

\lstdefinelanguage{Jac}{
    sensitive=false,
    morekeywords={
        sem,import,include,
        node,edge,walker,obj,
        enum,architype,
        has,can,glob,
        entry,exit,activity,
        spawn,take,disengage,report,
        anchor,
        if,elif,else,
        for,while,with,by,in,to,
        and,or,not, visit,
        continue,break,skip,
        try,except,finally,raise,
        assert,yield,return,
        from,as,
        true,false,null,
        self,super,
        class,def,
        async,await,
        test,
        match,case,
        ignore,context,info,details,
        strict,length,
        digraph,subgraph,
        Method,Function,
        py,jac
    },
    morecomment=[l]{//},
    morecomment=[l]{\#},
    morestring=[b]",
    morestring=[b]',
    morestring=[b]`
}

\lstdefinestyle{myjac}{
    language=Jac,
    basicstyle=\ttfamily\scriptsize,
    keywordstyle=\color{keyword}\bfseries,
    stringstyle=\color{string},
    commentstyle=\itshape\color{comment},
    identifierstyle=\color{identifier},
    numbers=left,
    numberstyle=\tiny\color{gray},
    stepnumber=1,
    numbersep=8pt,
    frame=single,
    rulecolor=\color{codeframe},
    tabsize=4,
    showstringspaces=false,
    showspaces=false,
    showtabs=false,
    breaklines=true,
    breakatwhitespace=false,
    keepspaces=true,
    columns=fullflexible,
    linewidth=\linewidth,
    xleftmargin=0.5cm,
    xrightmargin=0.1cm,
    breakindent=0pt,
    captionpos=b,
    emph={
        int,float,str,bool,list,dict,tuple,set,
        type,Any,
    },
    emphstyle=\color{emphcolor}\bfseries,
    postbreak={}
}

\lstnewenvironment{jac}[1][]{
    \lstset{style=myjac,#1}
}{}

\lstdefinestyle{myoutput}{
    backgroundcolor=\color{outbg},
    basicstyle=\ttfamily\footnotesize\color{outtext},
    numbers=none,
    frame=single,
    rulecolor=\color{outframe},
    showstringspaces=false,
    showspaces=false,
    showtabs=false,
    breaklines=true,
    breakatwhitespace=false,
    keepspaces=true,
    columns=flexible,
    captionpos=b,
    xleftmargin=0.5cm
}

\lstnewenvironment{out}[1][]{
    \lstset{style=myoutput,#1}
}{}

%% file: sections/abstract.tex
\begin{abstract}
Building agentic applications requires combining large language model (LLM) operations, workflow structure and application state, and model-guided routing. With SDKs and agent frameworks, developers connect these components through supporting code that prepares model inputs, propagates state, and dispatches model-selected steps. Workflow changes can require coordinated updates to this code, increasing development and testing effort. Splitting execution logic between application code and framework internals also limits compiler visibility into control flow and state dependencies, hindering automatic optimization. We introduce \textbf{Agent-Native Programming}, which unifies these three components with shared execution semantics. We realize this approach in \textbf{\sysname{}}, a Python-superset language, through \texttt{by llm()} for typed model operations and tool use, Object-Spatial Programming for workflow structure and state, and \texttt{visit ... by llm()} for model-guided routing. The routing construct derives the model's choices from executable workflow steps and schedules the selected steps. This explicit connection between model decisions and program execution gives the compiler and runtime a common basis for dependency analysis and automatic optimization. Across five benchmarks and five baselines, \sysname{} achieves a $1.8$--$8.7\times$ reduction in AI-integration code relative to the OpenAI SDK, with competitive observed task quality and lower coding-agent development effort. Two case studies demonstrate speculative execution and service partitioning through configuration alone, using program analysis and runtime checks to preserve workflow behavior under stated conditions.
\end{abstract}

%% file: sections/intro.tex
\section{Introduction}
\label{sec:intro}

Agentic applications use large language models (LLMs) together with external tools to carry out tasks through a sequence of model decisions and program actions~\cite{yao2023react}. For example, CodeAgent is an LLM-based coding assistant that investigates and repairs software issues. It plans an investigation, explores repository files, edits code, and verifies the result. After verification, the model uses the task goal and the information gathered so far to decide whether to edit again or finish.

Building such an application requires three kinds of support. \emph{Model operations} prepare model inputs, specify expected outputs and available tools, and extract the requested results from model responses. For CodeAgent, these operations provide the issue description and relevant code to the LLM and obtain outputs such as a plan or proposed code changes. \emph{Workflow structure and application state} describe the steps of a task, the paths between them, and the information retained during execution. The steps and paths form a task graph. A workflow can also specify which steps run in parallel and where to spawn sub-agents for separate tasks. In CodeAgent, the workflow connects planning, file exploration, editing, and verification, while application state records the information gathered and the results of each action. For example, such a workflow could spawn sub-agents to inspect different repository files in parallel and combine their findings before editing. \emph{Model-guided routing (orchestration)} uses a model during execution to choose the next step from the alternatives available in the task graph, based on the task goal and current application state. In CodeAgent, this determines whether to make further edits or finish the task after verification.

\begin{figure}[t]
    \centering
    \includegraphics[width=\linewidth]{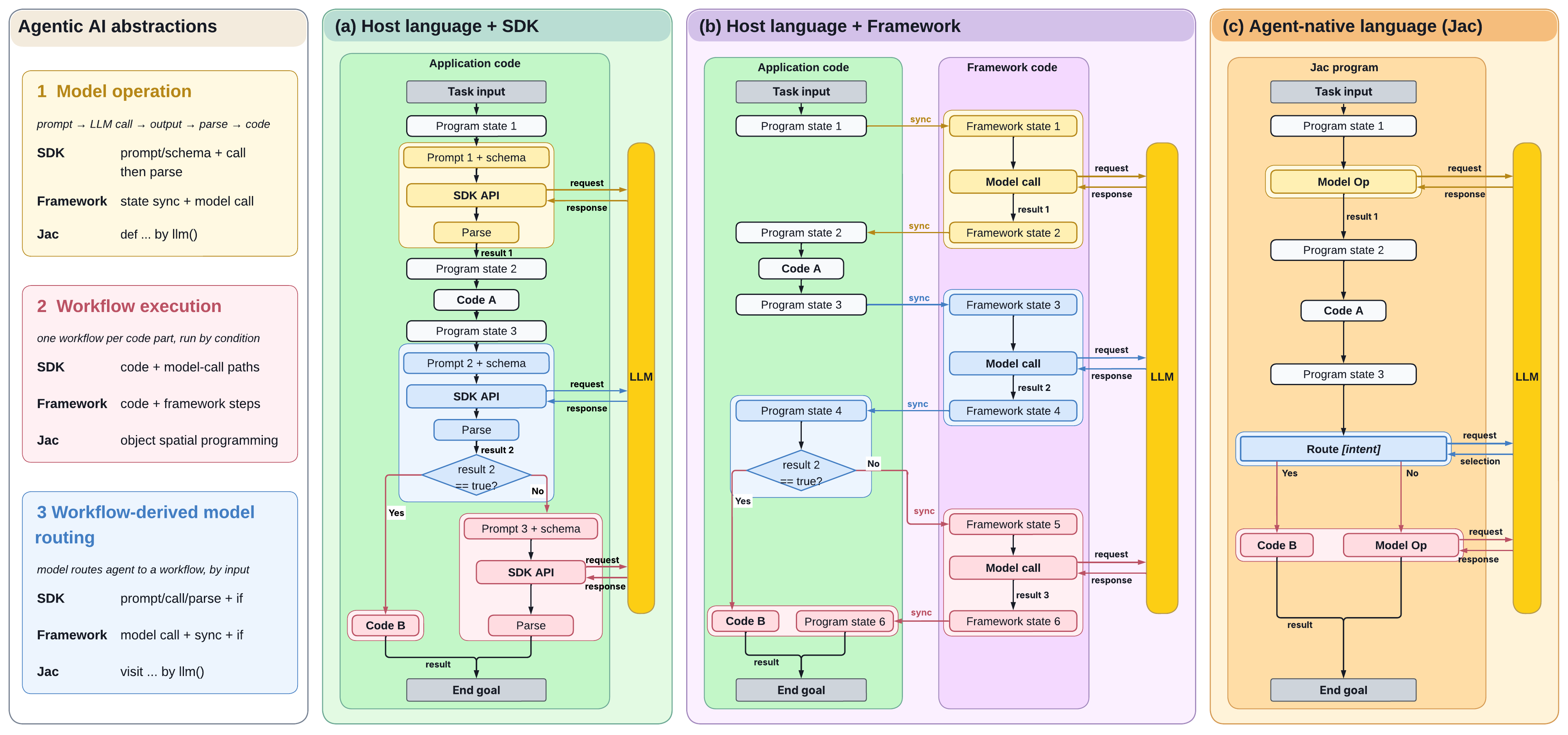}
    \caption{An agentic workflow fragment expressed through three programming approaches. (a) Application code builds model inputs, updates state, and runs the selected steps. (b) Frameworks handle parts of model interaction and workflow execution, while application code connects their interfaces to workflow steps and state. (c) \sysname{} expresses model operations, workflow structure and application state, and model-guided routing in one language, making their connections explicit to the compiler and runtime.}
    \label{fig:intro}
\end{figure}

Implementing these agentic applications in a general-purpose language involves invoking models through APIs for reasoning and workflow decisions, often using an SDK such as the OpenAI SDK~\cite{openai_python_sdk}. The SDK sends requests to the model and returns responses to the application. Application code must implement workflow steps and supporting code that prepares inputs for SDK calls, passes results between steps, and runs the steps selected by the model (Figure~\ref{fig:intro}(a)). Changes to model operations, workflow structure, state representations, or model-guided routing (orchestration) can therefore require corresponding updates to the supporting code. Keeping these updates consistent and testing them together adds development effort; missed updates can select the wrong step or omit required data.

Agent frameworks such as LangGraph~\cite{langgraph}, CrewAI~\cite{crewai}, and DSPy~\cite{khattab2024dspy} offer an alternative by reducing supporting code through reusable model and workflow abstractions (Table~\ref{tab:magic-table-wrap-features-fixed}). However, application code must still supply model inputs, update application state from returned results, and specify how model decisions select workflow steps (Figure~\ref{fig:intro}(b)). This requires synchronizing application and framework state and writing supporting code to run the steps selected by the model. Changes to workflow steps or state can therefore require corresponding updates to this supporting code, adding development and testing effort. Also, as execution depends on both application code and framework behavior, neither alone gives the compiler and runtime holistic visibility: a unified view of model operations, possible next steps, and the data each step reads or changes~\cite{lin2024parrot,zheng2024sglang}. The compiler and runtime must recover this view through additional analysis before they can safely apply automatic optimizations, such as starting a possible next step's model request before routing finishes.

These limitations motivate a system that reduces development effort while giving the compiler and runtime a shared view of model operations, workflow steps, and application state. Our central research question is: \emph{How can we simplify the construction of agentic workflows while exposing the information needed for automatic optimization?}

\textbf{Our key insight is that expressing model operations, workflow structure and application state, and model-guided routing in a unified language makes the connection between model decisions and program execution explicit. The language defines which steps a model can select, how selected steps are scheduled, and how they access application state.} The same workflow structure identifies both the model's choices and the steps the runtime can execute. This gives the compiler and runtime a common execution model for analyzing dependencies and checking when automatic optimizations apply.

Based on this insight, we present \textbf{Agent-Native Programming}, a programming model that brings these three parts together. The program specifies model inputs and expected outputs, workflow steps and state, and the steps available at each routing decision. The language handles model interactions and defines how selected steps execute. We realize this model in \sysname{}, a Python-superset programming language, as illustrated in Figure~\ref{fig:intro}(c).

\begin{enumerate}
    \item \emph{Model operations.}
    \sysname{} builds on Meaning-Typed Programming (MTP) through \texttt{by llm()}~\cite{dantanarayana2025mtp}. We extend MTP to tool-using agents and define the execution semantics of model operations with tool calls. The byLLM runtime constructs requests from typed declarations and input values. It executes model-requested tool calls and returns their results to the model before converting the final response to the declared output type.

    \item \emph{Workflow structure and application state.}
    \sysname{} uses Object-Spatial Programming (OSP), which extends object-oriented programming with explicit graph structure~\cite{mars2025osp}. We define its execution semantics for agentic workflows. Nodes represent workflow steps and local data, while edges represent possible transitions. A walker represents one run of the workflow and carries state across steps. Entry code on the node or walker implements each step using node data and walker state. The \texttt{visit} statement schedules steps. Walkers spawned to delegate work execute as sub-agents and can also run in parallel.

    \item \emph{Model-guided routing.}
    \sysname{} supports \emph{workflow-derived model routing} through \texttt{visit ... by llm()}. This construct derives the model's available choices from workflow steps identified at runtime by a graph expression and schedules the selected steps. Application code therefore does not need to maintain a separate mapping from model responses to workflow steps.
\end{enumerate}

We evaluate \sysname{} on five agentic application benchmarks against the OpenAI SDK, LangGraph, CrewAI, DSPy, and NOOA~\cite{nooa}. In expert-written implementations, \sysname{} achieves a $1.8$--$8.7\times$ reduction in AI-integration code relative to the SDK, with competitive observed task quality across three model families. We also measure development effort using Claude Code and Codex as coding-agent \emph{builders}. We count their tokens during implementation, testing, and revision, and check whether the resulting applications meet shared quality requirements. When builders choose the application design, \sysname{} builds consume $48\%$ fewer tokens with Claude Code and $51\%$ fewer with Codex than SDK builds, aggregated across the five applications.

We also demonstrate compiler and runtime support for automatic optimization through two case studies. Both are enabled through application configuration without changes to application code. They use program analysis and runtime checks and rely on the stated conditions for preserving workflow behavior. Relative to the same \sysname{} applications without speculation, speculative execution yields median routing-decision latency savings of $29\%$--$39\%$ on 3 of the 5 benchmarks. For service partitioning, which separates application components into services, the baselines consume $3.8$--$8.7\times$ as many builder tokens as \sysname{}, with all evaluated partitions passing the required correctness checks.

This paper makes the following contributions:
\begin{itemize}
    \item \textbf{Programming model and semantics.}
    We realize Agent-Native Programming in \sysname{} with common
    execution semantics for model operations, workflow structure and
    application state, and model-guided routing. The language uses the workflow structure to determine which steps the model can select next and schedules the selected steps.

   \item \textbf{Compiler and runtime support.} We expose the relationships between model operations, workflow steps, model-guided routing, and application state, giving the compiler and runtime a holistic view for analyzing dependencies and applying automatic transformations.

    \item \textbf{Empirical evaluation.}
    Across five benchmarks, we demonstrate reduced AI-integration code
    and coding-agent development effort with competitive task quality,
    and quantify the benefits of both transformations.
\end{itemize}

%% file: sections/background.tex
\section{Motivation}
\label{sec:motivation}

\begin{figure}[ht]
    \centering
    \includegraphics[width=0.75\linewidth]{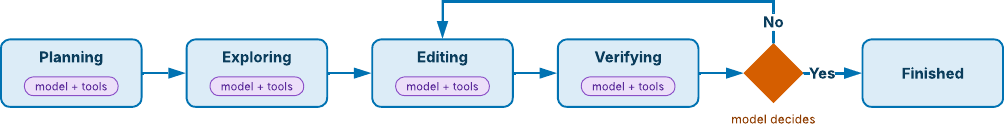}
    \caption{CodeAgent's workflow. The model performs the work within each workflow step with access to permitted tools and, after verification, decides whether to edit again or finish.}
    \label{fig:motivation-workflow}
    \vspace{-0.5cm}
\end{figure}

We use CodeAgent to examine the supporting code required for model operations, workflow structure and application state, and model-guided routing with the OpenAI SDK and LangChain with LangGraph. We then present the same application in \sysname{}, showing how language constructs give these elements common execution semantics and expose their relationships to the compiler and runtime.

\subsection{The CodeAgent Workflow}
\label{sec:motivation:example}

\begin{figure}
    \centering
    \includegraphics[width=0.8\linewidth]{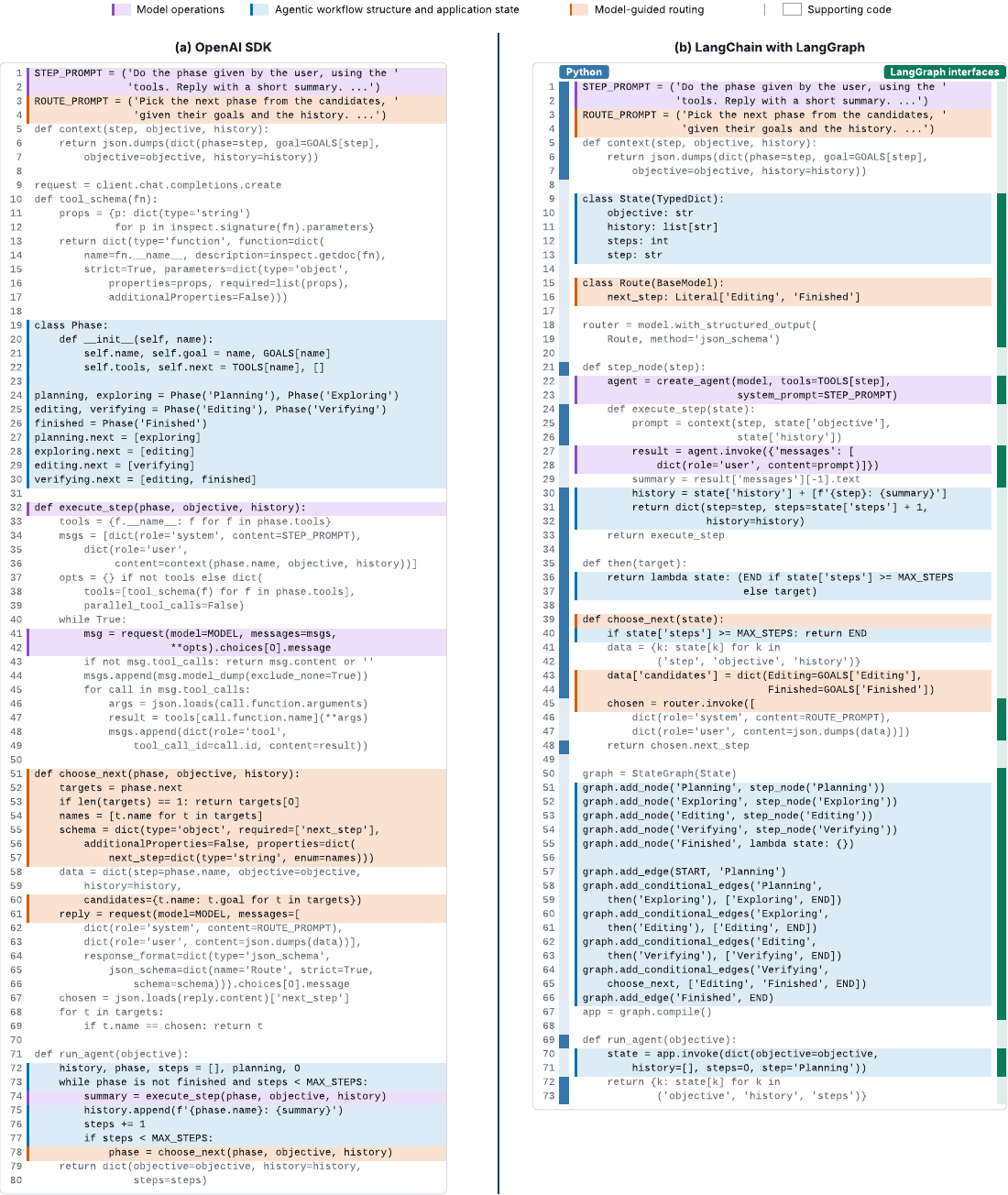}
    \caption{CodeAgent implemented with (a) the OpenAI SDK and (b) LangChain with LangGraph. Purple marks model operations, blue marks agentic workflow structure and application state, and orange marks model-guided routing. Unshaded regions show supporting code, including request construction, response interpretation, and tool or phase invocation. The sidebars in (b) identify application code and uses of LangGraph interfaces.}
    \label{fig:motivation-agent-comparison}
      \vspace{-0.5cm}
\end{figure}

Figure~\ref{fig:motivation-workflow} shows CodeAgent's workflow for investigating and repairing a software issue. We refer to its workflow steps as \emph{phases}. \texttt{Planning} outlines how to investigate and repair the issue, \texttt{Exploring} gathers evidence from repository files, \texttt{Editing} applies a repair, and \texttt{Verifying} runs tests to assess the result. Model operations carry out each phase's work with access to its permitted tools, while the application maintains the objective, a history of phase summaries, and the number of phase executions. After verification, a model uses the updated history to select Editing or \texttt{Finished}. If a regression test fails and Editing is selected, the application runs that phase again with access to the recorded test results.

Figures~\ref{fig:motivation-agent-comparison} and~\ref{fig:motivation-jac} show code sketches of this application using the OpenAI SDK, LangChain with LangGraph, and \sysname{}. All three implementations follow the same phases and transitions, use the same phase goals (\texttt{GOALS}) and permitted tools (\texttt{TOOLS}), and enforce the same limit on phase executions (\texttt{MAX\_STEPS}). We focus on how each implementation executes the agentic workflow by invoking models, maintaining application state, and using model decisions to determine which phase runs next.

\subsection{Agentic Programming with a Model SDK}
\label{sec:motivation:sdk}

Figure~\ref{fig:motivation-agent-comparison}(a) shows CodeAgent implemented with the OpenAI SDK~\cite{openai_python_sdk}. The \texttt{Phase} class defines phase names, goals, permitted tools, and successor links (lines~19--22), while lines~24--30 instantiate and connect the phases. These objects describe the agentic workflow, but application code must still implement model interactions, execute requested tools, and translate model decisions into transitions between phases.

\paragraph{Executing a phase.}
The application must implement a model--tool interaction loop in addition to the repository tools themselves. The \texttt{context} function serializes the phase goal, objective, and history into model input (lines~5--7), while \texttt{tool\_schema} constructs schemas for permitted tools (lines~10--17). Within this loop, \texttt{execute\_step} parses each requested tool's arguments, resolves its name to a callable, invokes it, and returns the result to the model (lines~45--49). The loop continues until the model produces a phase summary.

\paragraph{Turning a model decision into a transition.}
Although \texttt{phase.next} already identifies Verifying's successor phases, application code must translate those objects into model choices and translate the answer back into an executable phase. The \texttt{choose\_next} function constructs the output schema and candidate descriptions (lines~52--60), then parses the returned name and resolves it to a successor object (lines~67--69) before \texttt{run\_agent} can advance the workflow (lines~73--78). The successor links therefore specify the available transitions, while application code implements how the model's choice becomes one of those transitions.

For the regression scenario, \texttt{run\_agent} must append Verifying's summary (line~75) before calling \texttt{choose\_next} (line~78), then supply the updated history to the selected phase on the next iteration (line~74). If that history is not supplied correctly, a valid phase selection can still lead to a repair that ignores the failed test. Application code therefore defines both how the model's answer selects the next phase and which state that phase receives.

\subsection{Agentic Programming with Frameworks}
\label{sec:motivation:frameworks}

\begin{wrapfigure}{r}{0.45\textwidth}
    \centering
    \includegraphics[width=\linewidth]{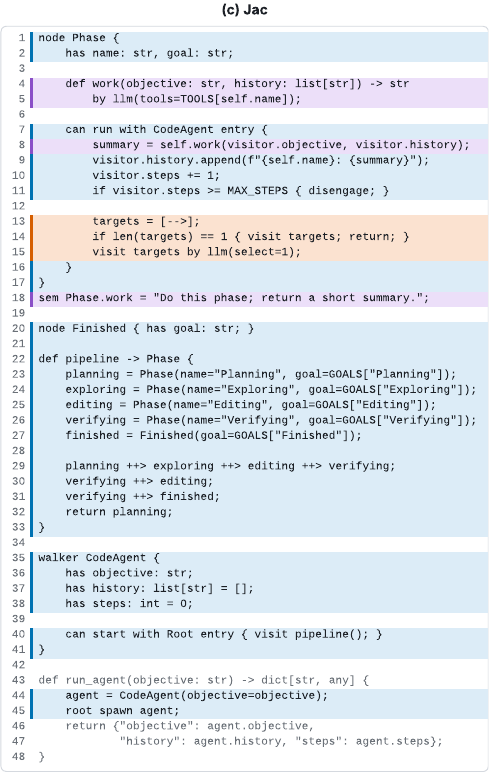}
    \caption{CodeAgent implemented in \sysname{}. Language constructs express model operations, workflow structure and application state, and workflow-derived model routing. Colors follow Figure~\ref{fig:motivation-agent-comparison}.}
    \label{fig:motivation-jac}
    \vspace{-0.3cm}
\end{wrapfigure}
Agent frameworks offer an alternative that reduces this work through reusable support for model interactions and workflow execution. Figure~\ref{fig:motivation-agent-comparison}(b) shows CodeAgent implemented with LangChain and LangGraph~\cite{langchain_agents,langgraph}, following the workflow in Figure~\ref{fig:motivation-workflow}. We use LangGraph to represent the phases and their possible transitions as graph nodes and edges (lines~50--66). The blue and green sidebars distinguish application code from uses of LangGraph interfaces, showing how the workflow spans both.

\paragraph{Executing a phase.}
LangChain's \texttt{create\_agent} supplies the model--tool loop (lines~22--23), but application code still connects phase inputs and results to workflow state. The callback created by \texttt{step\_node} constructs model input from the objective and history in LangGraph's state (lines~24--28). It extracts the summary from returned messages and packages the updated history and phase execution count as a state update (lines~29--32). This callback implements the call/result and state bindings in Figure~\ref{fig:intro}(b). These conversions determine what each phase's model receives and which results reach later model decisions. Changes to phase inputs, result formats, or state fields can therefore require revising and testing this supporting code.

\paragraph{Turning a model decision into a transition.}

In this implementation, Verifying's successors, \texttt{Editing} and \texttt{Finished}, are declared separately in three places. They appear in the \texttt{Route.next\_step} output type (lines~15--16), the candidate descriptions sent to the model (lines~43--44), and the destinations for Verifying's conditional edges (lines~64--65). The \texttt{choose\_next} function obtains the model's selection and returns its name to LangGraph (lines~45--48). Application code must keep the model's permitted answers and their descriptions consistent with the destinations that the framework can execute.

Suppose we extend the workflow so that the model can also select Exploring after a failed verification. Adding Exploring to the graph destinations (line~65) does not add it to the model's output type (line~16) or candidate descriptions (lines~43--44). All three declarations must change together to make the new transition available to the model and executable by the framework. A helper could generate these declarations, but the application would still need to define how graph destinations become model choices and how the selected answer advances the workflow.

Building such agentic applications therefore requires additional development effort to implement, test, and maintain these callbacks and routing declarations alongside the phase logic. Automatic transformations also require the compiler and runtime to relate each selected step to its code, model requests, and state accesses.

\subsection{CodeAgent in \sysname{}}
\label{sec:motivation:gap}
\label{sec:motivation:vision}

These examples motivate treating a model-guided transition as a language operation. The program identifies executable destinations and supplies application state, while the language defines how a model's selection continues execution. Figure~\ref{fig:motivation-jac} shows CodeAgent implemented in \sysname{}, with model operations, agentic workflow structure and application state, and workflow-derived model routing expressed through language constructs.

\paragraph{Model operations.}
A model operation accepts function arguments and returns a result of its declared output type. It can also use application-supplied tools to retrieve information or perform actions. In the purple regions, \texttt{work} declares its inputs, summary output, and permitted tools, using \texttt{by llm(...)} to delegate execution to a model (lines~4--5). The language constructs requests, executes model-requested tool calls, returns tool results to the model, and converts the model's final response into the declared output type. A phase can therefore obtain its summary by calling \texttt{work} (line~8), without implementing the model--tool loop or response conversions.

\paragraph{Agentic workflow structure and application state.}
Workflow structure and application state are defined together with the code that executes each phase. In the blue regions, phases and successor links form the workflow (lines~23--31), while the objective, history, and number of phase executions are stored as application state (lines~36--38). Phase code updates this state directly (lines~9--10), making its results available to later phases without packaging them for a framework's state interface.

\paragraph{Workflow-derived model routing.}
Workflow-derived model routing selects among executable destinations identified by the program. The orange region illustrates this through \texttt{visit targets by llm(select=1)} (line~15), using the successors collected in \texttt{targets} (line~13). After Verifying, selecting Editing continues execution there with access to the updated history. The language gives the selection this execution meaning, without application code converting a returned phase name into a destination.

Under this model, adding \texttt{verifying ++> exploring} alongside the existing successor declarations (lines~30--31) makes Exploring both an executable successor and an option for the model at the existing decision point. This single declaration replaces the coordinated changes to graph destinations, the output type, and candidate descriptions required in the framework sketch.

\paragraph{Enabling automatic optimization.}
Expressing model operations, routing, and state access together gives the compiler and runtime holistic visibility into the workflow. For CodeAgent, this connects \texttt{Editing} to the code that prepares its model request and the state that code reads. This information supports dependency analysis and automatic transformations under checked conditions, as described in \S~\ref{sec:compiler-runtime}. The next section defines the language support and execution semantics for this programming model.

%% file: sections/design.tex
\section{Agent-Native Programming in Jac}
\label{sec:anp}
\label{sec:anp-core}

We realize Agent-Native Programming in Jac through language constructs for model operations, workflow structure and application state, and model-guided orchestration. These constructs define how model and tool calls execute, how workflow steps access application state, and how the runtime schedules steps selected by the model. We use the CodeAgent implementation in Figure~\ref{fig:motivation-jac} to illustrate their execution semantics.

\subsection{Model Operations and Tools}
\label{sec:model-operations}

In Jac, we define a model operation as a typed function or method whose body is replaced by a
\texttt{by} clause.

\paragraph{Declaring a model operation.}
In CodeAgent, each phase does its work through the method \texttt{work} of the
node type \texttt{Phase} (Figure~\ref{fig:motivation-jac}, lines 1--5 and 18):
\begin{jac}
node Phase { # type of a CodeAgent phase
    has name: str, goal: str;
    # typed inputs and output
    def work(objective: str, history: list[str]) -> str
        by llm(tools=TOOLS[self.name]); # model and permitted tools
}
# intent of Phase.work
sem Phase.work = "Do this phase; return a short summary.";
\end{jac}
In general, a model operation has the form
$\mathbf{def}\ f(p_1:T_1,\ldots,p_n:T_n)\ \to\ T_r\ \mathbf{by}\ m(\mathbf{tools}=\mathcal F)$,
where $m$ is a model, $T_r$ is a supported result type, and $\mathcal F$ is the set of tools available to the model. A \texttt{sem} declaration adds task intent not explicitly expressed by
the function or method signature. It identifies the target by its
qualified name, here \texttt{Phase.work}.

\paragraph{Invoking the operation.}
A model operation is called like any other function or method (Figure~\ref{fig:motivation-jac}, line 8):
\begin{jac}
# self: a Phase node
summary = self.work(visitor.objective, visitor.history);
\end{jac}
This call asks the model to perform the current phase's task using
the workflow objective and history ($\bar v$). The current phase,
\texttt{self}, supplies its name and goal as context.
\texttt{TOOLS[self.name]} selects the tools $\mathcal F$ available
to that phase. The byLLM runtime combines these inputs with the function signature,
the \texttt{sem} string, and descriptions of the available tools to
build the model request. It maintains a history $\mu$ of tool requests
and their results for this invocation, initially $\mu=\langle\rangle$.

The runtime sends the prepared request, including history $\mu$,
to the model and receives a response $R=\operatorname{ask}(\mu)$. The model either provides an answer or requests a tool.
An answer is represented as $\operatorname{final}(\hat v)$, where
$\hat v$ is the model's output before conversion to the declared
return type. A tool request is represented as
$\operatorname{call}(g,\hat a)$, where $g$ is the requested tool
and $\hat a$ contains arguments supplied by the model.

\paragraph{Handling the response.}
Once the model responds, the runtime either converts the final answer or executes a requested tool and asks the model again. We describe this process by $\operatorname{run}(\mu)$, with one response $R=\operatorname{ask}(\mu)$ obtained at each iteration:
\begin{equation}
\operatorname{run}(\mu)=
\begin{cases}
\operatorname{convert}_{T_r}(\hat v),
  &R=\operatorname{final}(\hat v),\\
\operatorname{run}(\mu\cdot\langle g,\hat a,z\rangle),
  &R=\operatorname{call}(g,\hat a),\ g\in\mathcal F.
\end{cases}
\label{eq:model-continuation}
\end{equation}

A final answer $\hat v$ is converted to the declared return type $T_r$, including any structure required by that type. If conversion succeeds, the operation returns the converted value. This ensures that the result has the declared type, but does not guarantee that the answer is correct.

For a tool request, the runtime converts the supplied arguments $\hat a$ to the parameter types of $g$ and executes the tool, producing result $z$. It appends the tool request and result to history $\mu$ before asking the model again. For example, in CodeAgent's \texttt{Verifying} phase, the model may request a test tool and use its output to produce a verification summary. The runtime converts this summary to \texttt{str} and returns it to the caller.

\subsection{Agentic Workflow Structure and State}
\label{sec:workflow-state}

In Jac, we use Object-Spatial Programming (OSP) to organize and execute stateful agentic workflows as graphs. We define application state, abilities, traversal, and sub-agent execution in terms of OSP constructs. A node stores step data, an edge represents a possible transition, and a walker carries state for one workflow invocation. The following linear CodeAgent example connects four \texttt{Phase} nodes, referenced by \texttt{planning} through \texttt{verifying}, and starts a walker. Figure~\ref{fig:motivation-jac} shows their creation (lines 23--26) and adds edges from \texttt{Verifying} back to \texttt{Editing} and on to \texttt{Finished} (lines 30--31).
\begin{jac}
node Phase { # continues Section 3.1
    can run with CodeAgent entry { # runs when a walker arrives
        summary = self.work(visitor.objective, visitor.history);
        # self: node, visitor: walker
        visitor.history.append(f"{self.name}: {summary}");
        visit [-->]; # schedule the successors
    }
}
walker CodeAgent {
    has objective: str, history: list[str] = []; # state of one run
    # self: walker, here: node
    can start with Root entry { visit [-->]; }
}
with entry {
    # connect the phases
    root ++> planning ++> exploring ++> editing ++> verifying;
    # start a walker at root
    root spawn CodeAgent(objective="Fix the failing test");
}
\end{jac}

\paragraph{State and abilities.}
Nodes and walkers store state in typed fields. In CodeAgent, each \texttt{Phase} node stores its name and goal, while a \texttt{CodeAgent} walker carries the objective and history of one workflow invocation. The \texttt{++>} operator connects the nodes with edges. An entry ability is code that the runtime executes automatically when a walker arrives at a node. Its form is $\mathbf{can}\ a\ \mathbf{with}\ T\ \mathbf{entry}\ \{s\}$, where $a$ names the ability, $T$ specifies the type that triggers it, and $s$ is its body. For an ability declared on a node, $T$ specifies the visiting walker type. For example, \texttt{Phase.run} executes when a \texttt{CodeAgent} walker arrives at the phase. Inside this ability, \texttt{self} refers to the phase and \texttt{visitor} to the walker, allowing the phase to read and update \texttt{visitor.history}. For an ability declared on a walker, $T$ specifies the visited node type. For example, \texttt{CodeAgent.start} executes when the walker enters \texttt{Root}. Inside this ability, \texttt{self} refers to the walker and \texttt{here} to the current node.

In the traversal rules below, $A(n,w)$ denotes the sequence of entry abilities executed when walker $w$ arrives at node $n$.

\paragraph{Traversal.}
A walker $w$ tracks its current node $\ell$ and a queue $Q$ of nodes to visit. A \texttt{visit} statement evaluates $e$ once and adds the resulting destinations $V$ to the queue:
\begin{equation}
 \mathbf{visit}\ e:\quad (\ell,Q)\longrightarrow(\ell,Q\cdot V),
 \quad\text{where }V=\operatorname{eval}(e),
 \label{eq:enqueue}
\end{equation}
Here $V$ is a finite sequence of node references, with a single node treated as a one-element sequence, and $\cdot$ appends sequences. The current ability continues after \texttt{visit}. Once the current node's entry abilities finish, the walker moves to the next queued node $n$ and runs $A(n,w)$, or ends if the queue is empty:
\begin{equation}
 \operatorname{advance}:\quad (\ell,n\cdot Q)\longrightarrow(n,Q)
 \text{ and run }A(n,w);\quad(\ell,\langle\rangle)\longrightarrow\operatorname{end},
 \label{eq:traversal-completion}
\end{equation}
A \texttt{disengage} statement stops the walker immediately, even if nodes remain queued (Figure~\ref{fig:motivation-jac}, line 11):
\begin{equation}
 \mathbf{disengage}:\quad (\ell,Q)\longrightarrow\operatorname{end}.
 \label{eq:disengage}
\end{equation}

In the CodeAgent example above, \texttt{visit [- ->]} at \texttt{Planning} queues \texttt{Exploring}. After Planning's abilities finish, the walker moves to Exploring and runs its entry abilities. Revisits preserve the walker's updated state, so returning from \texttt{Verifying} to \texttt{Editing} makes the latest verification result available to Editing.

\paragraph{Sub-agents.}
A workflow delegates a task by starting a separate walker, called a sub-agent:
\begin{jac}
# start the reviewer
task = flow root spawn Reviewer(change=diff);
review = wait task; # wait for the reviewer to finish
\end{jac}
Here, \texttt{flow} starts \texttt{Reviewer} at \texttt{root} and immediately returns a task handle. The next statement, \texttt{wait task}, pauses the calling code until the reviewer finishes, then returns the completed walker as \texttt{review}. Other work can be placed between these statements to execute while the reviewer runs. Using \texttt{spawn} without \texttt{flow} waits for the walker to finish before executing the next statement. Each walker has its own location, queue, and fields, while referenced objects may be shared.

\subsection{Model-Guided Orchestration}
\label{sec:model-guided-orchestration}

In Jac, we use model-guided orchestration to let a model choose which workflow nodes a walker visits next. The program supplies candidate nodes, and \texttt{visit ... by llm} asks the model to select among them. CodeAgent uses this construct to choose its next phase (Figure~\ref{fig:motivation-jac}, lines 13 and 15):
\begin{jac}
targets = [-->]; # collect the current node's successors
visit targets by llm(select=1); # select and schedule at most one
\end{jac}
Adding an edge from \texttt{Verifying} to \texttt{Exploring} makes \texttt{Exploring} another candidate at that point.

\paragraph{Selecting and scheduling workflow steps.}
The runtime evaluates the argument $e$ of \texttt{visit} once to obtain the finite candidate list $C$. It describes each candidate using the node's type and fields, such as its name and goal, and sends these descriptions with the current node and walker state to the model. Each candidate has a label, and the model returns the labels of its choices.

The runtime maps these labels back to nodes in $C$, ignores labels that identify no candidate, and includes each node only once. It collects nodes in response order up to a limit of $k$, where $k$ is a positive integer. The resulting sequence $V$ is appended to the visit queue:
\begin{equation}
\mathbf{visit}\ e\ \mathbf{by}\ m(\mathbf{select}=k):\ 
(\ell,Q)\longrightarrow(\ell,Q\cdot V),
\quad\operatorname{set}(V)\subseteq\operatorname{set}(C),\quad |V|\leq k.
\label{eq:model-guided-visit}
\end{equation}

As in Equation~(\ref{eq:enqueue}), this adds nodes to $Q$ without changing the current node $\ell$. The walker later visits them through $\operatorname{advance}$ and runs their entry abilities. Jac handles the mapping from model choices to nodes, so the application needs no separate routing code.

For example, after CodeAgent's \texttt{Verifying} phase in Figure~\ref{fig:motivation-jac}, $C=\langle\texttt{Editing},\texttt{Finished}\rangle$. Choosing \texttt{Editing} gives $V=\langle\texttt{Editing}\rangle$, which queues that node for execution. If $C$ is empty, no model request is sent; if $V$ is empty, no nodes are added.

\paragraph{What the runtime guarantees.}
Equations~(\ref{eq:model-continuation}) and~(\ref{eq:model-guided-visit}) define the limits enforced by the runtime.

Proposition 3.1 (Runtime containment).
A successful model operation returns a value of its declared type $T_r$. The runtime executes a tool requested by the model only if it belongs to $\mathcal F$. A successful model-guided visit schedules at most $k$ nodes from $C$, with no duplicates in that selection. These limits follow from converting answers, checking requested tools against $\mathcal F$, and filtering and limiting selected nodes. 

Together, Equations~(\ref{eq:model-continuation})--(\ref{eq:model-guided-visit}) provide a common execution semantics for model operations, workflow execution and application state, and model-guided orchestration.

%% file: sections/compiler_runtime.tex
\section{Compiler and Runtime Support}
\label{sec:compiler-runtime}

\begin{figure}[t]
  \centering
  \includegraphics[width=\linewidth]{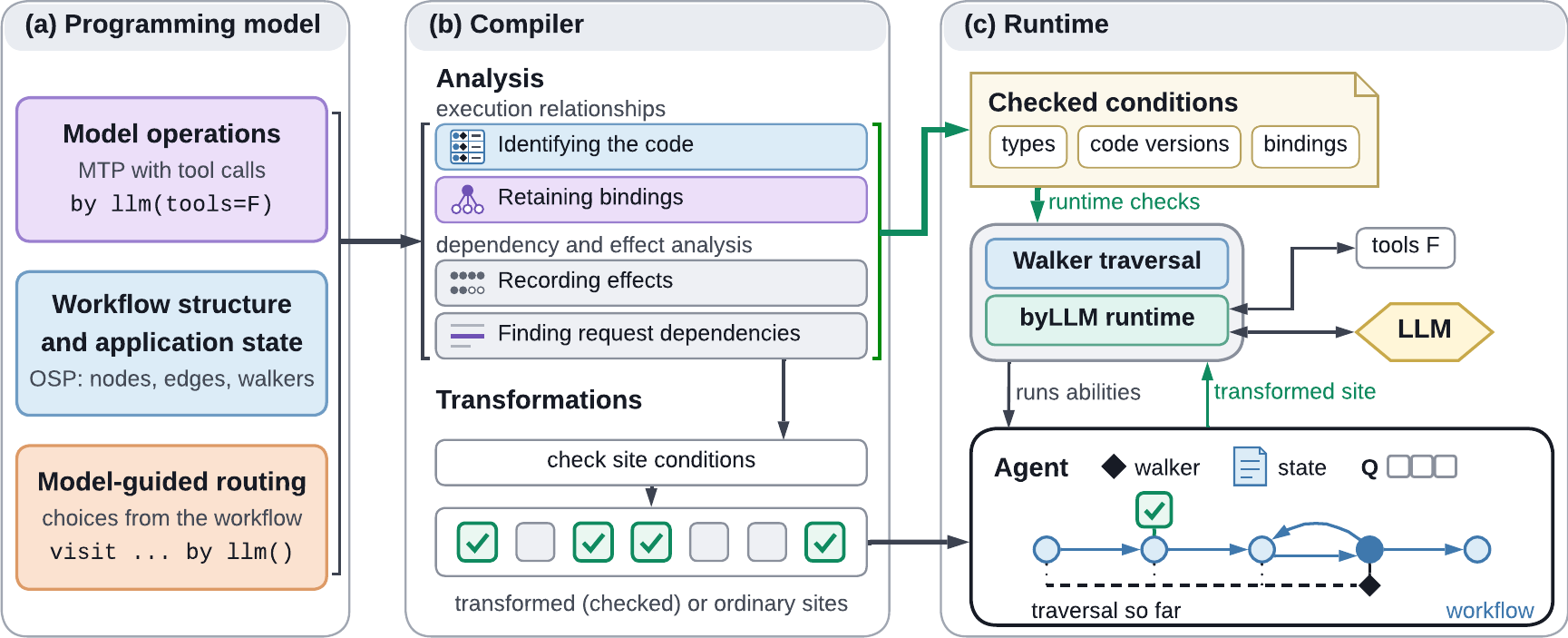}
  \Description{The compiler connects model declarations, ability bodies, and their state dependencies. The runtime supplies concrete nodes, walkers, and tool bindings and executes the resulting code.}
  \caption{Compiler and runtime architecture. The compiler identifies possible code, state accesses, and model-request inputs. The runtime supplies concrete objects and values. Transformations use this information to change execution under checked conditions.}
  \label{fig:compiler-pipeline}
\end{figure}

By defining how model operations, workflow steps, and model-guided routing execute together (Figure~\ref{fig:compiler-pipeline}(a)), Jac connects each possible next step to its code, model requests, and state. The compiler analyzes dependencies through these connections, while the runtime supplies each run's node and walker objects and values (Figure~\ref{fig:compiler-pipeline}(b)--(c)). This shared view supports automatic optimizations under checked conditions. Using CodeAgent (Figure~\ref{fig:motivation-jac}), we illustrate this compiler and runtime support with speculative model requests, one possible optimization.

\subsection{Execution Relationships and Runtime Bindings}
\label{sec:execution-structure}
\label{sec:lowering}
\label{sec:runtime-coordination}
\label{sec:context-binding}

\paragraph{Identifying the code.} In CodeAgent, an Editing node has type \texttt{Phase}. Its \texttt{run} ability calls \texttt{self.work} with the CodeAgent walker's \texttt{objective} and \texttt{history}. The type on which an ability is declared and the type that triggers it determine whether it belongs to $A(n,w)$, the entry abilities for node $n$ and walker $w$ defined in \S~\ref{sec:anp}. We write $\mathcal{A}(T_c,T_w)$ for the abilities compatible with node type $T_c$ and walker type $T_w$, from which $A(n,w)$ is drawn. The graph's concrete nodes and edges are runtime state, so the compiler identifies compatible code from types, while the runtime supplies the objects used by each invocation.

\paragraph{Retaining bindings.} The compiler tracks the node and walker bindings defined in \S~\ref{sec:anp} through calls. For \texttt{work}, it retains declaration metadata and the expressions supplying its receiver, arguments, and configuration. Each tool in $\mathcal F$ retains its callable and, for a method, its bound receiver. We call the compatible abilities and these retained bindings the \emph{agent execution graph}. Generated code preserves these connections and where execution continues after each call. At a guided visit, the runtime evaluates the candidate expression once to obtain $C$, retains its node references during selection, and appends the selected nodes $V$ to the queue $Q$ shown in Figure~\ref{fig:compiler-pipeline}(c). When a node is reached, its abilities use the current objects. Model requests similarly use evaluated receiver, argument, and configuration values. These objects may share state, which is why \S~\ref{sec:dependency-effects} treats aliases conservatively.

\subsection{Dependency and Effect Analysis}
\label{sec:dependency-effects}

\paragraph{Recording effects.} For each body or slice $b$, the effect analysis in Figure~\ref{fig:compiler-pipeline}(b) records
\begin{equation}
  \operatorname{Eff}(b)=
  \langle \operatorname{Rd}(b),\operatorname{Wr}(b),
          \operatorname{Ext}(b),\operatorname{Req}(b),
          \operatorname{Sched}(b)\rangle .
  \label{eq:agent-effects}
\end{equation}
The first two components are sets of locations that may be read or written; the others are flags for external application interactions, model requests, and scheduling. Writes to fresh, non-escaping local storage are excluded from $\operatorname{Wr}$. Locations cover fields, globals, and resources; possible aliases overlap conservatively. The compiler combines direct effects with callee summaries through receiver and argument bindings. Summaries come from analyzed bodies or \emph{trusted declarations} of callee effects, whose correctness is assumed; missing summaries yield unknown effects. Monotone updates over a finite location abstraction reach a fixed point.

\paragraph{Finding request dependencies.} Backward slicing follows value and control dependencies and possible reaching writes, including through aliases. The request slice includes receiver and argument evaluation, configuration, rendering, tool selection, and schema construction. Node and walker fields provide initial bindings, but code may access other state. When unresolved calls or dependencies prevent establishing a transformation's conditions, the program executes without that transformation.

For Editing's first \texttt{work} request, the slice $S$ has read footprint $\operatorname{Rd}(S)$: the phase's \texttt{name} and \texttt{goal}, the walker's \texttt{objective} and \texttt{history}, and \texttt{TOOLS[self.name]}. This lookup selects tools without executing them. Here $\operatorname{Wr}(S)$ is empty and the three flags are false. The compiler analyzes request preparation separately from tool execution and the history update in Figure~\ref{fig:motivation-jac}, line 9.

\subsection{Example: Speculative Model Requests}
\label{sec:speculative-requests}
\label{sec:execution-continuation}

\begin{wrapfloat}{algorithm}{r}{0.48\columnwidth}
\footnotesize
\setlength{\emergencystretch}{1em}
\caption{Checking candidate eligibility}
\label{alg:agent-compilation}
\begin{algorithmic}[1]
\REQUIRE Typed program $P$; candidate and walker types $T_c,T_w$.
\ENSURE Request site $r$, slice $S$, read footprint, and bindings $B$, or rejection.
\STATE $\mathcal A\gets\operatorname{ResolveEntries}(P,T_c,T_w)$.
\STATE Reject unresolved entry abilities.
\STATE $\mathcal R\gets\operatorname{FirstModelSites}(P,\mathcal A)$.
\STATE Reject unresolved sites or $|\mathcal R|\neq1$.
\STATE $r\gets$ the unique site in $\mathcal R$.
\STATE Reject unless every entry path reaches $r$ as its first model request.
\STATE $S\gets\operatorname{RequestSlice}(P,\mathcal A,r)$.
\STATE Reject unresolved dependencies.
\STATE Compute $\operatorname{Eff}(S)$ from bodies and declared summaries.
\STATE Reject unless $S$ has known effects and bounded execution.
\STATE Reject if $\operatorname{Wr}(S)\neq\emptyset$.
\STATE Reject if $\operatorname{Ext}(S)$ or $\operatorname{Req}(S)$ or $\operatorname{Sched}(S)$.
\STATE $B\gets$ bindings required by $S$.
\RETURN $(r,S,\operatorname{Rd}(S),B)$.
\end{algorithmic}
\end{wrapfloat}
Speculative model requests illustrate how the preceding analysis and runtime checks work together to optimize execution. After Verifying, CodeAgent (Figure~\ref{fig:motivation-jac}) asks the model to select from $C=\langle\text{Editing},\text{Finished}\rangle$. An auxiliary predictor proposes a candidate from $C$ while the routing model makes its decision. For an eligible prediction such as Editing, the runtime can issue the first \texttt{work} request early. The routing decision and reuse checks determine whether the response is used. Only the first $\mathit{ask}(\mu)$, with $\mu=\langle\rangle$, is issued early, because later requests depend on tool results. Tool execution and history updates remain part of normal execution.

Speculation requires compiler and runtime checks. The compiler verifies that request preparation has no application effects; the runtime verifies that a response matches the request at the original call.

\paragraph{Checking candidate eligibility.} Algorithm~\ref{alg:agent-compilation} checks candidate and walker types at the ``check site conditions'' step in Figure~\ref{fig:compiler-pipeline}(b). Every entry path must reach the same first model request; its preparation must have resolved dependencies, bounded execution, and no application effects.

For an eligible site, the compiler emits preparation code, its read footprint $\operatorname{Rd}(S)$, the required bindings $B$, and a guarded response lookup (the \emph{request guard}) at the original call. Ineligible sites execute without speculation.

Preparation executes the slice, excluding the request itself, rather than the candidate's entry prefix. It reads captured inputs, writes private storage, and calls only analyzed or summarized helpers. An update that affects the request or its bindings enters the slice and prevents this read-only preparation; an unrelated one remains in normal execution. In Figure~\ref{fig:motivation-jac}, \texttt{work} is the first statement of \texttt{run} (line 8), so $S$ has no entry-prefix code; Finished has no model operation, so predicting it launches nothing.

\paragraph{Launching the request.} For a predicted candidate in $C$, the runtime checks types, code versions, and bindings against the analysis, as shown in the ``Checked conditions'' box in Figure~\ref{fig:compiler-pipeline}(c). It captures the slice's read footprint consistently using synchronization or version validation, then releases source-state access before sending. Capture failures or preparation exceptions cancel the launch. The raw response remains private; even a request for a tool $g\in\mathcal F$ executes nothing yet. Its slot identifies the routing occurrence, candidate, and target model site, so repeated visits remain distinct; auxiliary jobs are cancellable.

\paragraph{Checking response reuse.} Selecting Editing appends it to $Q$. The rest of the routing ability and other walkers may still change state, so reuse compares the complete request instead of trusting the earlier snapshot. At the original call, normal execution constructs the current request. Reuse requires a ready, unconsumed slot for this occurrence, an identical complete request (model, settings, messages, schemas, and response format), and matching callable and receiver identities. The runtime consumes a matching response once; otherwise, it sends the current request normally, without waiting for unfinished speculation. Both paths resume the response handling in Equation~(\ref{eq:model-continuation}). Preparation changes no application state, so entry updates are neither replayed nor rolled back; a misprediction discards at most one speculative request.

\paragraph{Preserving workflow behavior.} The eligibility check in Algorithm~\ref{alg:agent-compilation} ensures that speculative preparation changes no application state. The runtime either consumes a response matching the complete request once or sends the request normally, then continues with the same response handling; unused responses remain private. This argument assumes sound dependency and effect summaries, including trusted declarations, and fair scheduling that prevents auxiliary work from delaying normal execution indefinitely. It also assumes that possible model responses depend only on the complete request and that extra requests change neither the outcomes nor the availability of normal requests. Under these assumptions, ignoring auxiliary requests and buffering, every optimized execution has application state changes, tool interactions, results, failures, and completion permitted by the original program. Bounded preparation, cancellable jobs, and fallback without waiting preserve progress. The claim concerns which behaviors are possible; response probabilities may differ.

%% file: sections/eval.tex
\section{Evaluation}
\label{sec:eval}

Our evaluation addresses the following research questions:

\begingroup
\renewcommand{\theenumi}{RQ\arabic{enumi}}
\renewcommand{\labelenumi}{\textbf{\theenumi.}}
\begin{enumerate}
\item \label{rq:development-complexity}\textbf{Development effort.} How do \sysname{}'s abstractions affect AI-integration code size and coding-agent token usage under open and fixed application designs?

\item \label{rq:task-quality}\textbf{Task quality.} How does the observed task quality of expert-written \sysname{} applications compare with existing approaches across application models?

\item \label{rq:execution-optimization}\textbf{Compiler and runtime benefits.} What benefits can the compiler and runtime derive from \sysname{}'s explicit representation of model computation, state, and traversal?
\end{enumerate}
\endgroup

\subsection{Experimental Setup}
\label{sec:eval:setup}

\subsubsection{Applications and Baselines}
\label{sec:eval:applications}
\label{sec:eval:baselines}
We evaluate Meeting~\cite{crewai_meeting_example}, Email~\cite{crewai_email_example}, RagGPT~\cite{jacgpt_application}, YTNavigator~\cite{ytnavigator_application}, and CodeAgent~\cite{jimenez2024swebench} against five baselines: OpenAI Python SDK~\cite{openai_python_sdk} with hand-written prompts, JSON schemas, and orchestration, LangGraph~\cite{langgraph}, DSPy~\cite{khattab2024dspy}, CrewAI~\cite{crewai}, and NOOA~\cite{nooa}. The expert implementations adapt the same application designs to each programming interface and share code for functionality unrelated to AI integration. These implementations are used for the code-size and task-quality studies; the coding-agent studies use independently constructed implementations. Table~\ref{tab:eval:benchmarks} describes the applications' origins, datasets, and scoring criteria.

\begin{table*}[!t]
    \centering
    \caption{Benchmark applications, their origins and evaluation datasets, and the item scores used for task quality. Dataset sizes count distinct inputs, before repetitions. For a 1--5 judge rating $r$, $J(r)=(r-1)/4$ maps the rating to $[0,1]$; $C$ denotes labeled key-point coverage. Higher scores are better.}
    \label{tab:eval:benchmarks}
    \small
    \setlength{\tabcolsep}{5pt}
    \renewcommand{\arraystretch}{1.13}
    \resizebox{\textwidth}{!}{%
    \input{figures/eval_benchmarks.tex}
    }
\end{table*}

\begin{figure*}[!t]
    \centering
    \begin{subfigure}[t]{.49\textwidth}
        \centering
        \includegraphics[width=\linewidth]{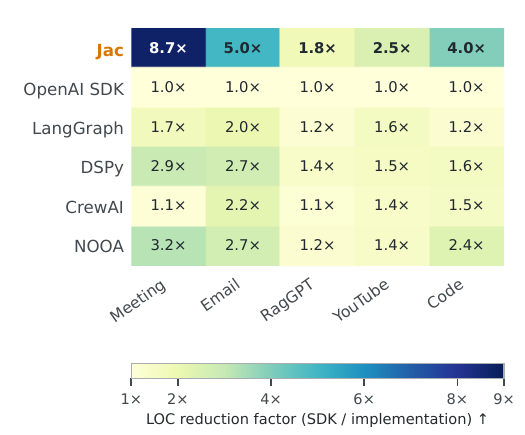}
        \caption{}\label{fig:eval-rq1-code-a}
    \end{subfigure}\hfill
    \begin{subfigure}[t]{.49\textwidth}
        \centering
        \includegraphics[width=\linewidth]{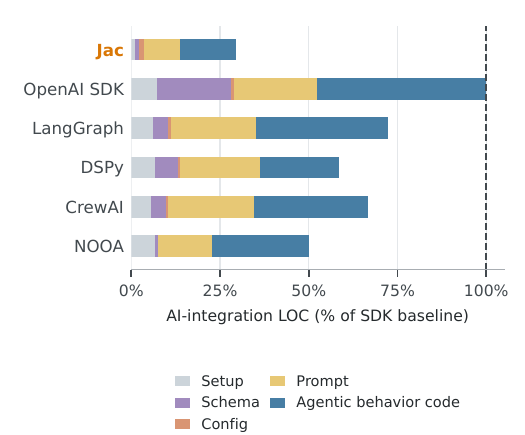}
        \caption{}\label{fig:eval-rq1-code-b}
    \end{subfigure}
    \caption{AI-integration code in expert implementations. (a) SDK LOC divided by implementation LOC for each application. (b) Category totals across five applications, normalized to total SDK LOC. Agentic behavior includes model handling and framework structure; \texttt{sem} descriptions count as prompt code.}
    \label{fig:eval:ai-code}
\end{figure*}

\subsubsection{Coding-Agent Builders and Models}
\label{sec:eval:protocol}
As developers increasingly delegate implementation to coding agents~\cite{bogdanov2026codingagents}, we assess development effort using Claude Code~\cite{eval_claude_code} and Codex~\cite{eval_codex_cli} as builders. We measure their token consumption during implementation, testing, and revision, and evaluate the resulting applications against shared quality requirements.

For each application, both builders receive the same task prompt, with skill files for the selected approach. We provide \sysname{} skills and framework-specific skill files obtained from official and community sources~\cite{eval_langgraph_skills,eval_crewai_skills,eval_nooa_skills,eval_dspy_skills_community}. Each builder constructs every application three times in \sysname{} and three times in each of the five baselines. Each build runs independently in a fresh autonomous session using the builder's default tools. We do not limit token usage, the number of turns, or build time.

We provide development tests for implementation and debugging. We evaluate completed applications on a separate test set that builders cannot access during construction. A build passes if it matches or exceeds the expert-written SDK implementation on every evaluated quality metric.

For construction, we use Claude Code \texttt{2.1.310} with Claude Opus 5.5~\cite{eval_opus55} and Codex CLI \texttt{0.167.0-alpha.4} with GPT-6.1~\cite{eval_openai_models}. We run the generated applications with GLM-5.3~\cite{eval_glm53} and evaluate their outputs using the success criterion above. RQ2 uses separate implementations written by the authors. We evaluate these expert-written applications with Claude Sonnet 5~\cite{eval_sonnet5}, GLM-5.3, and Gemma 4 31B~\cite{eval_gemma4}. For RQ3, we used Claude Code version \texttt{2.1.310}, the same version used in RQ1.

\input{sections/eval-rq1}
\input{sections/eval-rq2}
\input{sections/eval-rq3}

%% file: figures/eval_benchmarks.tex
\begin{tabular}
{@{}p{.13\textwidth}p{.63\textwidth}p{\dimexpr.54\textwidth-4\tabcolsep\relax}@{}}
\toprule
\textbf{Benchmark} & \textbf{Task and dataset} & \textbf{Scoring} \\
\midrule
\textbf{Meeting} &
\textbullet{} \textbf{Task/source:} Extract tasks and file them to mock Trello, CSV, and Slack outputs; adapted from CrewAI's Meeting Assistant Flow~\citep{crewai_meeting_example}.\par\smallskip
\textbullet{} \textbf{Dataset:} 100 labeled transcripts covering ordinary meetings, duplicate or retracted tasks, no-action cases, and prompt injection. &
$(F_1+C)/2$: semantic task precision/recall and key-point coverage. Missed tasks receive zero coverage. No-action transcripts score 1 for no predicted tasks, otherwise 0. \\
\addlinespace[4pt]
\textbf{Email} &
\textbullet{} \textbf{Task/source:} Select emails requiring replies and draft responses using thread context and mock search; adapted from the CrewAI--LangGraph example~\citep{crewai_email_example}.\par\smallskip
\textbullet{} \textbf{Dataset:} Synthetic inbox batches containing 290 emails, including 130 requiring replies, with empty-inbox and prompt-injection cases. &
Correctly skipping a reply scores 1; missed or unnecessary replies score 0. Required replies receive $(C+J(r))/2$ for key-point coverage and readiness to send. Routed and agentic results are averaged. \\
\addlinespace[4pt]
\textbf{RagGPT} &
\textbullet{} \textbf{Task/source:} Route requests among five specialists sharing a documentation index; adapted from Jac-GPT~\citep{jacgpt_application}.\par\smallskip
\textbullet{} \textbf{Dataset:} 220 items (240 turns) covering documentation, coding, debugging, small-talk, off-topic requests, and multi-turn context. &
Documentation, coding, debugging, and referenced multi-turn openings use $J(r)$. Other turns use binary appropriateness. \\
\addlinespace[4pt]
\textbf{YTNavigator} &
\textbullet{} \textbf{Task/source:} Answer channel questions using semantic search and SQL; adapted from YT-Navigator~\citep{ytnavigator_application}.\par\smallskip
\textbullet{} \textbf{Dataset:} 230 labeled question-answer pairs over a synthetic channel with 10 videos and 68 transcript chunks, covering factual answers, tool use, refusal, and multi-turn context. &
Normalized judge score $J(r)$ for factual correctness and completeness against reference answers. \\
\addlinespace[4pt]
\textbf{CodeAgent} &
\textbullet{} \textbf{Task/source:} Repair repository issues through planning, exploration, editing, and testing; evaluated on SWE-bench~\citep{jimenez2024swebench}.\par\smallskip
\textbullet{} \textbf{Dataset:} 360 pinned instances with identical issue descriptions, starting commits, and test harnesses across implementations. &
Resolution rate: an instance scores 1 if all \texttt{FAIL\_TO\_PASS} tests pass and all \texttt{PASS\_TO\_PASS} tests remain passing; otherwise 0. No LLM judge. \\
\bottomrule
\end{tabular}

%% file: sections/eval-rq1.tex
\subsection{\ref{rq:development-complexity}: Development Effort}
\label{sec:eval:rq1}

\begin{figure*}[t]
    \centering
    \includegraphics[width=\textwidth]{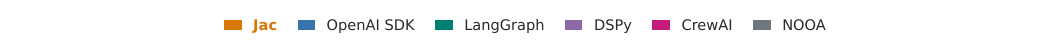}\par
    \begin{subfigure}[t]{.243\textwidth}
        \centering
        \includegraphics[width=\linewidth]{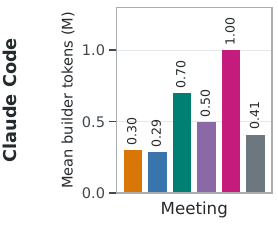}
    \end{subfigure}\hfill
    \begin{subfigure}[t]{.184\textwidth}
        \centering
        \includegraphics[width=\linewidth]{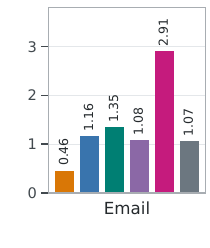}
    \end{subfigure}\hfill
    \begin{subfigure}[t]{.184\textwidth}
        \centering
        \includegraphics[width=\linewidth]{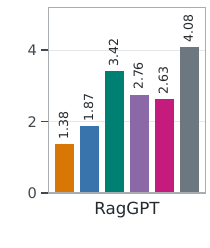}
    \end{subfigure}\hfill
    \begin{subfigure}[t]{.184\textwidth}
        \centering
        \includegraphics[width=\linewidth]{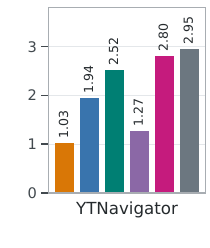}
    \end{subfigure}\hfill
    \begin{subfigure}[t]{.184\textwidth}
        \centering
        \includegraphics[width=\linewidth]{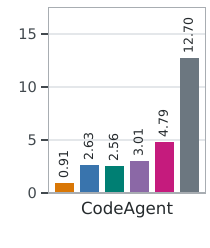}
    \end{subfigure}
    \par\medskip
    \begin{subfigure}[t]{.243\textwidth}
        \centering
        \includegraphics[width=\linewidth]{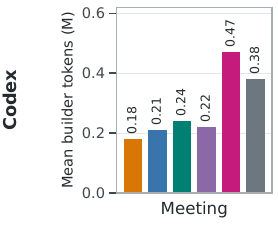}
    \end{subfigure}\hfill
    \begin{subfigure}[t]{.184\textwidth}
        \centering
        \includegraphics[width=\linewidth]{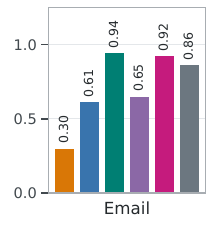}
    \end{subfigure}\hfill
    \begin{subfigure}[t]{.184\textwidth}
        \centering
        \includegraphics[width=\linewidth]{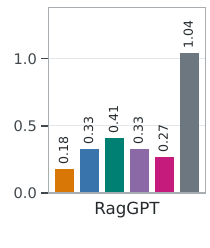}
    \end{subfigure}\hfill
    \begin{subfigure}[t]{.184\textwidth}
        \centering
        \includegraphics[width=\linewidth]{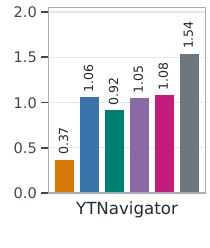}
    \end{subfigure}\hfill
    \begin{subfigure}[t]{.184\textwidth}
        \centering
        \includegraphics[width=\linewidth]{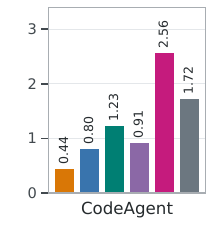}
    \end{subfigure}
    \caption{Building agentic applications from requirements with Claude Code using Claude Opus 5.5 (top) and Codex using GPT-6.1 (bottom). Bars show mean builder tokens over three independent builds per application in \sysname{} and each baseline. Counts include input, output (including reasoning), and cache reads/writes.}
    \label{fig:eval:open-design}
\end{figure*}

\subsubsection{AI-Integration LOC Comparison}
\label{sec:eval:ai-code}
We compare the lines of code (LOC) needed to implement the agentic functionality of each benchmark application. Prior work on language model programming systems also evaluates implementation effort using LOC~\cite{beurerkellner2023lmql,dantanarayana2025mtp}. AI-integration LOC counts additions relative to an application with model calls replaced by stubs, including prompts, schemas, setup, configuration, and agentic behavior. We exclude existing application logic and code reused unchanged across implementations for functionality unrelated to AI integration. We count natural-language \texttt{sem} descriptions in \sysname{} as prompt code.

Figure~\ref{fig:eval:ai-code}(a) shows that every baseline requires more AI-integration LOC than \sysname{} in all five applications. The SDK requires $1.8$--$8.7\times$ the LOC of \sysname{}. LangGraph requires approximately $1.5$--$5.1\times$ the LOC of \sysname{}, DSPy $1.3$--$3.0\times$, CrewAI $1.6$--$7.9\times$, and NOOA $1.5$--$2.7\times$.

Figure~\ref{fig:eval:ai-code}(b) compares integration LOC summed across the five applications. The SDK requires $3.38\times$ the LOC of \sysname{}. Even after excluding prompt lines from both the SDK and \sysname{} implementations, the SDK requires $3.90\times$ the LOC of \sysname{}. Schema definitions and agentic-behavior code account for approximately 73\% of \sysname{}'s LOC savings relative to the SDK. \sysname{} also uses less prompt and agentic-behavior code than each of the four framework baselines.

\subsubsection{Building Agentic Applications from Requirements}
\label{sec:eval:open-design}
\label{sec:eval:builder-work}
In the open-design setting, we give the coding agents the application requirements and success criteria. Each builder then determines the architecture and control flow needed to satisfy those requirements.

We measure development effort using the total tokens a coding agent consumes during implementation, testing, and revision until the application satisfies the given requirements. We count input, output, and cache-read/write tokens; output tokens include reasoning tokens. For each coding agent and application, we report the mean total token count over three independent builds in \sysname{} and in each baseline.

\begin{figure*}[t]
    \centering
    \includegraphics[width=\textwidth]{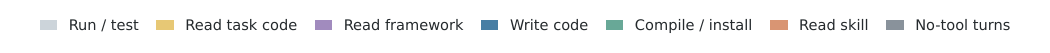}\par
    \begin{subfigure}[t]{.48\textwidth}
        \centering
        \includegraphics[width=0.9\linewidth]{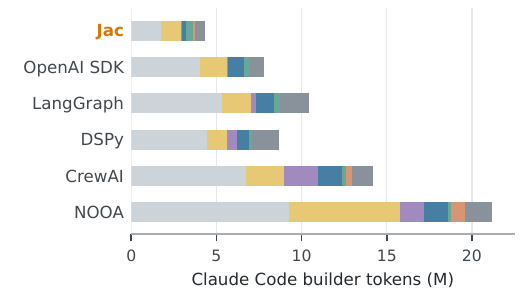}
        \caption{}\label{fig:eval:builder-work-claude}
    \end{subfigure}\hfill
    \begin{subfigure}[t]{.48\textwidth}
        \centering
        \includegraphics[width=0.9\linewidth]{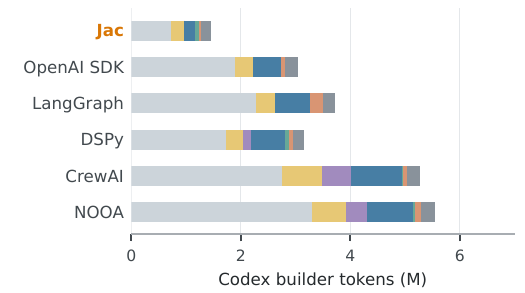}
        \caption{}\label{fig:eval:builder-work-codex}
    \end{subfigure}
    \caption{Mean builder tokens by activity over three independent open-design builds, summed across five applications: (a) Claude Code; (b) Codex. Counts use the same builds as Figure~\ref{fig:eval:open-design}. We count all tokens from each builder interaction, including reused context, under the activity performed. Uncategorized Codex shell commands count as run/test.}
    \label{fig:eval:builder-work}
\end{figure*}

Figure~\ref{fig:eval:open-design} shows that both builders use fewer tokens with \sysname{} than with LangGraph, DSPy, CrewAI, and NOOA on all five applications. Across the five applications, Claude Code uses approximately 53--81\% fewer tokens with \sysname{} than with these frameworks, and Codex uses approximately 53--73\% fewer. Relative to the SDK, the reductions are 48\% and 51\%, respectively. Claude Code uses the fewest tokens with \sysname{} on four of the five applications, and Codex does so on all five.

Figure~\ref{fig:eval:builder-work} shows mean token use by activity over three independent open-design builds, summed across five applications. We count all tokens from each builder interaction under its activity; for example, an interaction that runs tests contributes to run/test. Claude Code uses 0.22M tokens writing \sysname{} code, compared with 0.73M--1.42M for LangGraph, DSPy, CrewAI, and NOOA, and 0.93M for the SDK. It uses 2.94M tokens reading, running, and testing \sysname{} applications, compared with 5.63M--15.76M with these frameworks and 5.63M with the SDK. Both builders use fewer tokens with \sysname{} in both measures than with every baseline, with larger savings from reading and testing than writing code.

\subsubsection{Building Agentic Applications from Fixed Designs}
\label{sec:eval:fixed-spec}

\begin{wrapfigure}{r}{0.46\textwidth}
\vspace{-0.35cm}
    \centering
    \includegraphics[width=\linewidth]{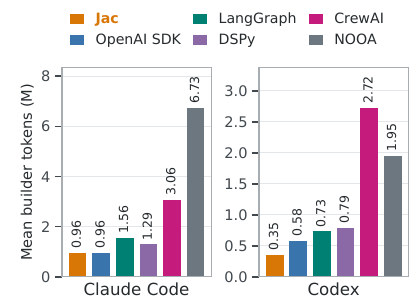}
    \caption{Building agentic applications from fixed designs with Claude Code and Codex. Bars show mean builder tokens per build across application designs.}
    \label{fig:eval:fixed-spec}
    \vspace{-0.2cm}
\end{wrapfigure}

In the fixed-design setting, we add a prescribed application architecture, control flow, tools, instructions, and execution budgets to the requirements in \S~\ref{sec:eval:open-design}. We otherwise use the same construction protocol.

Even when we prescribe the architecture and control flow, both builders use fewer tokens with \sysname{} than with LangGraph, DSPy, CrewAI, and NOOA (Figure~\ref{fig:eval:fixed-spec}). Claude Code averages 0.96M tokens per build with \sysname{} and the SDK, versus 1.29M--6.73M with these frameworks. Codex averages 0.35M with \sysname{}, versus 0.73M--2.72M with these frameworks and 0.58M with the SDK.

\textbf{RQ1 takeaway. \sysname{} reduces AI-integration code across the five applications and lowers aggregate builder token usage relative to the framework baselines in both design settings, while matching or reducing the SDK's token usage. Together, these results reflect lower development effort.}

%% file: sections/eval-rq2.tex
\subsection{\ref{rq:task-quality}: Task Quality}
\label{sec:eval:rq2}
\label{sec:eval:model-families}
\label{sec:eval:application-quality}
We compare task quality across \sysname{} and the five baselines using expert-written implementations of the benchmark applications. We evaluate each implementation using the application-specific metrics and success criteria in Table~\ref{tab:eval:benchmarks}.

For Meeting, Email, RagGPT, and YTNavigator, we use GPT-4.1-mini~\cite{eval_gpt41mini} as an LLM judge to assess whether application outputs meet the evaluation criteria, using application-specific references and rubrics. For CodeAgent, we follow SWE-bench's evaluation protocol~\cite{jimenez2024swebench} and report the resolution rate.

\begin{figure*}[!t]
    \centering
    \includegraphics[width=\textwidth]{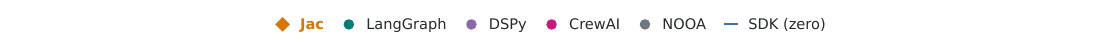}\par
    \begin{subfigure}[t]{.325\textwidth}
        \centering
        \includegraphics[width=\linewidth]{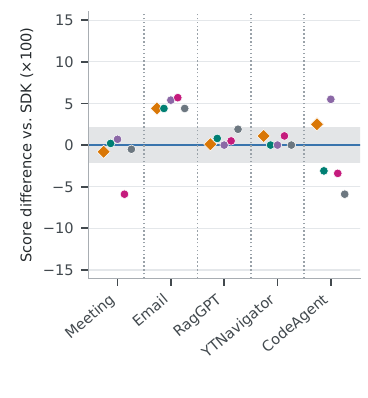}
        \caption{GLM-5.3}\label{fig:eval:quality-differences-a}
    \end{subfigure}\hfill
    \begin{subfigure}[t]{.325\textwidth}
        \centering
        \includegraphics[width=\linewidth]{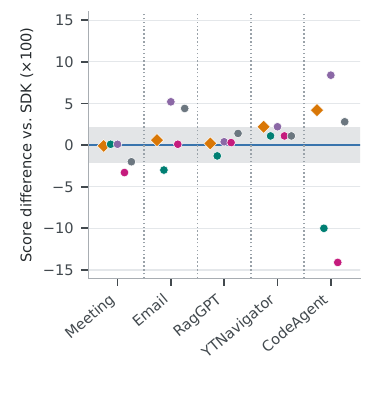}
        \caption{Claude Sonnet 5}\label{fig:eval:quality-differences-b}
    \end{subfigure}\hfill
    \begin{subfigure}[t]{.325\textwidth}
        \centering
        \includegraphics[width=\linewidth]{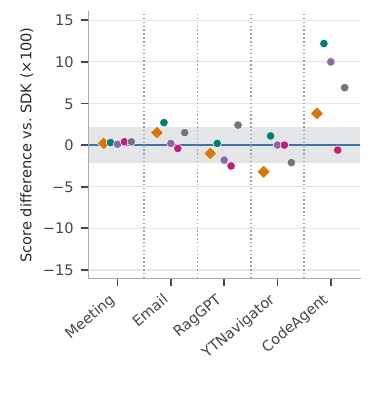}
        \caption{Gemma 4 31B}\label{fig:eval:quality-differences-c}
    \end{subfigure}
    \caption{Task quality relative to the SDK for matched application/model pairs: $100(s_{\mathrm{impl}}-s_{\mathrm{SDK}})$. Positive values favor the implementation. The gray $\pm2$-point band is a visual reference, not a confidence interval or equivalence threshold.}
    \label{fig:eval:quality-differences}
    \vspace{-0.6cm}
\end{figure*}

Across the 15 application/model pairs in Figure~\ref{fig:eval:quality-differences}, \sysname{}'s score differences from the SDK range from $-3.2$ to $+4.4$ points. This range is narrower than those of LangGraph, DSPy, CrewAI, and NOOA. The contrast is clearest on CodeAgent, where \sysname{}'s margin remains positive across all three models and changes by about 1.7 points, compared with 22.2 for LangGraph and 13.5 for CrewAI.

\textbf{RQ2 takeaway. Using \sysname{}, we can build agentic AI applications with less development effort while maintaining generally similar or better task quality than the baselines across the evaluated tasks.}

%% file: sections/eval-rq3.tex
\subsection{\ref{rq:execution-optimization}: Compiler and Runtime Benefits}
\label{sec:eval:rq3}

We conducted two case studies examining how Jac's holistic view of model operations, workflow steps, and application code supports reusable transformations. Workflow structure identifies possible next steps; their code determines the required data and effects. Jac exposes both through constructs and analysis in \S\S~\ref{sec:anp} and~\ref{sec:compiler-runtime}. Starting from working applications, we compare per-application work of enabling Jac's existing mechanisms with implementing each transformation separately in the baselines. We measure applicability, correctness under the evaluation checks, builder effort, and speculation latency. Builder effort excludes shared compiler and runtime development.

\subsubsection{Case Study 1: Speculative Execution}
\label{sec:eval:runtime-optimization}

Speculative execution reduces sequential latency by issuing a predicted next model request while the routing call runs~\cite{ye2026speculativeactions}. Safe overlap requires identifying the request and checking its preparation, since discarding a response cannot undo application effects. We evaluate the mechanism in \S~\ref{sec:speculative-requests} using Jev~\texttt{1.14}, TypeSafe AI's System One model for structured judgments from supplied state~\cite{almeida2026jev}, as the auxiliary predictor. Jev proposes a candidate while GLM-5.3 makes the routing decision. The compiler checks whether the request can be prepared early; runtime guards determine whether the response can be reused at the selected call.

\begin{figure*}[!t]
    \centering
    \includegraphics[width=.92\textwidth]{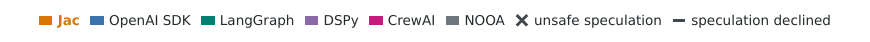}\\[2pt]
    \begin{subfigure}[t]{.325\textwidth}
        \centering
        \includegraphics[width=\linewidth]{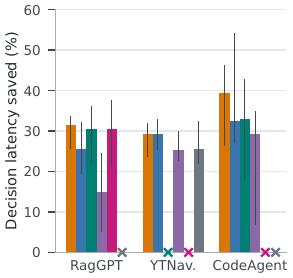}
        \caption{Latency saved}\label{fig:eval:spec-latency}
    \end{subfigure}\hfill
    \begin{subfigure}[t]{.325\textwidth}
        \centering
        \includegraphics[width=\linewidth]{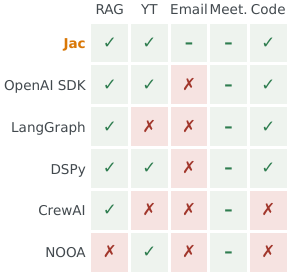}
        \caption{Correctness}\label{fig:eval:spec-correctness}
    \end{subfigure}\hfill
    \begin{subfigure}[t]{.325\textwidth}
        \centering
        \includegraphics[width=\linewidth]{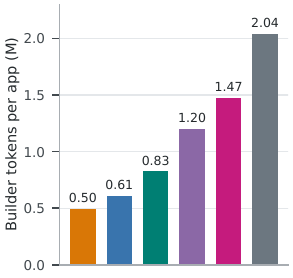}
        \caption{Builder tokens}\label{fig:eval:spec-tokens}
    \end{subfigure}
    \caption{Speculative execution across five applications. (a) Median paired routing-decision latency reduction, with 95\% bootstrap intervals; unsafe implementations are excluded. (b) Safe speedup ($\checkmark$), correct refusal (--), or unsafe speculation ($\times$). (c) Mean builder tokens, including refusals and unsafe builds.}
    \label{fig:eval:spec-builds}
\vspace{-0.5cm}
\end{figure*}

\paragraph{Automating the safety decision.}
With an SDK, a compiler must recover model operations and workflow transitions from ordinary calls and application code. Frameworks expose more workflow structure, but the effects of preparing requests still depend on the code inside their steps. LangGraph, for example, represents state and routing explicitly, while its nodes are Python functions that can invoke models or other code~\cite{langgraphGraphApi}. A compiler for Python and LangGraph could perform this analysis, but must interpret framework dispatch, follow application and library calls, and combine their dependencies and effects. The workflow graph alone does not establish these effects. Jac supplies these connections through language constructs, without requiring the compiler to recover their meaning from framework APIs.

Each framework/application pair receives one autonomous coding-agent session with the task, safety requirements, and framework-specific instructions. Jac's instructions explain how to enable its existing mechanism through configuration without changing application code. Its compiler applies deterministic eligibility rules for a given program and set of summaries, declines uncertain cases, and generates runtime checks under the assumptions in \S~\ref{sec:speculative-requests}. The evaluated baseline setups provide no corresponding compiler pass, so builders infer applicability from code and implement speculation and its checks through available APIs. Jac places these decisions in reusable analysis rather than a new model judgment for each application. Builders may self-test on supplied tests; separate held-out checks classify builds as safe speculation, correct refusal, or unsafe speculation under the evaluation's rules. Only safe implementations receive latency credit.

\paragraph{Correct speculation includes correct refusal.}
Jac accelerates RagGPT, YTNavigator, and CodeAgent while passing the held-out checks, and correctly declines on Meeting and Email (Figure~\ref{fig:eval:spec-correctness}). Email's next request depends on information retrieved after routing, so early preparation requires an action forbidden before selection. All baseline builds for Meeting also decline, but all five baseline Email builds violate the evaluated rules by performing prohibited work or speculating outside the permitted continuation. Failures on eligible workloads bring the total to 10 of 25 unsafe baseline builds. In these builds, Jac's shared mechanism produces safe speculation or a correct refusal for every application.

\paragraph{Latency benefits depend on the portion of execution overlapped.}
We measure routing-decision latency from the routing call's start to the end of the selected handler's first model response, and express paired savings relative to this interval without speculation. Jac saves a median 31\% on RagGPT, 29\% on YTNavigator, and 39\% on CodeAgent, while safe baseline implementations obtain savings of 15--33\% (Figure~\ref{fig:eval:spec-latency}).

\paragraph{Lower builder effort accompanies correct applicability decisions.}
Jac uses a mean of 0.50M builder tokens per application, compared with 0.61M--2.04M for the baselines (Figure~\ref{fig:eval:spec-tokens}). These means include refusals and unsafe builds. On Email, baseline builders spend 0.68M--1.74M tokens yet all produce unsafe implementations, while Jac spends 0.71M and correctly declines.

\subsubsection{Case Study 2: Automatic Service Partitioning}
\label{sec:eval:deployment}

Service partitioning connects execution structure to application code and state. Function signatures identify arguments and results, but not the workflow state and resources that must remain with a service. Service Weaver shows how declared component interfaces support generated communication~\cite{ghemawat2023moderncloud}; we examine whether Jac can use existing application declarations. A walker carries workflow state across steps, and its node encounters connect that state to abilities, model operations, and tools. These relationships give the compiler a unit of execution beyond a function call and, with callable bindings and dependency analysis, support interface and ownership checks. SDK programs can express this organization, but ordinary function signatures do not supply these workflow relationships.

\begin{wrapfigure}{r}{0.48\textwidth}
\vspace{-0.5cm}
    \centering
    \includegraphics[width=\linewidth]{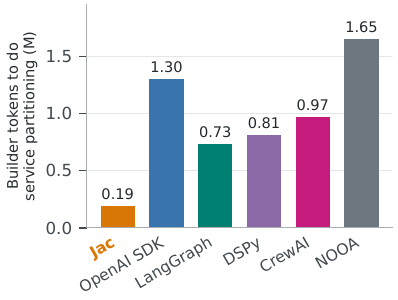}
    \caption{Mean builder tokens to partition an existing HTTP application into prescribed services, averaged across five applications. All evaluated partitions pass the deployment and correctness checks.}
    \label{fig:eval:deployment}
    \vspace{-0.5cm}
\end{wrapfigure}

Starting from an existing HTTP application, each builder creates the prescribed agent and dependency services while preserving instructions, tools, and control flow. Jac's builder configures the partitioner and checks the deployment without changing application code; the compiler generates client and service bindings from existing declarations. Baseline builders implement wrappers, clients, and deployment configuration. We supply the service boundaries and measure the work of implementing them. Every evaluated partition passes the deployment and correctness checks: the checker uses a client with an unreachable model endpoint and checks service roles and model-call attribution to verify that tested executions use the intended services.

Jac uses a mean of 0.19M builder tokens per application, compared with 0.73M--1.65M for the baselines, or approximately $3.8$--$8.7\times$ Jac's effort (Figure~\ref{fig:eval:deployment}). All evaluated partitions pass the checks; enabling Jac's partitioner uses fewer builder tokens on average than implementing service interfaces and communication code in the baselines.

\textbf{RQ3 takeaway. \sysname{}'s holistic view of model operations, workflow structure, application code and state, and model-guided routing enables the compiler and runtime to automate speculation and service partitioning while checking the conditions for their safe application.}

%% file: sections/related.tex
\section{Related Work}
\label{sec:related-work}

\begin{wraptable}[15]{r}{0.47\textwidth}
\vspace{-1cm}
\centering
\caption{Support for agentic application abstractions.
  \protect\ding{51}: full support;
  $\sim$: partial support requiring integration code;
  $\times$: no dedicated support.}
\label{tab:magic-table-wrap-features-fixed}

\scriptsize
\setlength{\tabcolsep}{1pt}
\setlength{\arrayrulewidth}{0.35pt}
\renewcommand{\arraystretch}{1.28}

\newcommand{\rotLabel}[2]{%
  \shortstack[c]{%
    \textbf{#1}\\[-1pt]
    {\tiny\citep{#2}}%
  }%
}

\newlength{\rotHeadHeight}
\newlength{\rotHeadMeasured}
\setlength{\rotHeadHeight}{0pt}

\newcommand{\measureRotLabel}[2]{%
  \settowidth{\rotHeadMeasured}{\rotLabel{#1}{#2}}%
  \ifdim\rotHeadMeasured>\rotHeadHeight
    \setlength{\rotHeadHeight}{\rotHeadMeasured}%
  \fi
}

\measureRotLabel{SDK}{openai_python_sdk}
\measureRotLabel{LangGraph}{langgraph}
\measureRotLabel{CrewAI}{crewai}
\measureRotLabel{NOOA}{nooa}
\measureRotLabel{DSPy}{khattab2024dspy}

\addtolength{\rotHeadHeight}{10pt}

\newcommand{\rotHeadCited}[2]{%
  \parbox[b][\rotHeadHeight][b]{\linewidth}{%
    \centering
    \rotatebox[origin=c]{90}{\rotLabel{#1}{#2}}%
    \par\vspace*{4pt}%
  }%
}

\newcommand{\rotHeadPlain}[1]{%
  \parbox[b][\rotHeadHeight][b]{\linewidth}{%
    \centering
    \rotatebox[origin=c]{90}{\strut\textbf{#1}}%
    \par\vspace*{4pt}%
  }%
}

\newcommand{\featureCell}[2]{%
  \raisebox{0pt}%
    [\dimexpr\height+4pt\relax]%
    [\dimexpr\depth+4pt\relax]{%
      \shortstack[l]{#1\\#2}%
    }%
}

\begin{tabular}{|
  >{\raggedright\arraybackslash}
    m{\dimexpr0.34\linewidth-2\tabcolsep-2\arrayrulewidth\relax}|
  *{6}{
    >{\centering\arraybackslash}
    m{\dimexpr0.11\linewidth-2\tabcolsep-\arrayrulewidth\relax}|
  }}
\hline

{} &
\rotHeadCited{SDK}{openai_python_sdk} &
\rotHeadCited{LangGraph}{langgraph} &
\rotHeadCited{CrewAI}{crewai} &
\rotHeadCited{NOOA}{nooa} &
\rotHeadCited{DSPy}{khattab2024dspy} &
\rotHeadPlain{Jac} \\
\hline

\featureCell{Declarative}{model operations} &
\magicNo & \magicSome & \magicSome &
\magicYes & \magicYes & \magicYes \\
\hline

\featureCell{Specified}{workflow execution} &
\magicNo & \magicYes & \magicYes &
\magicNo & \magicNo & \magicYes \\
\hline

\featureCell{Workflow-derived}{model routing} &
\magicNo & \magicSome & \magicSome &
\magicNo & \magicNo & \magicYes \\
\hline

\featureCell{Application-state}{propagation} &
\magicNo & \magicSome &
\magicYes & \magicYes & \magicSome & \magicYes \\
\hline

\end{tabular}
\vspace{-0.3cm}
\end{wraptable}

Table~\ref{tab:magic-table-wrap-features-fixed} compares the abstractions provided by the evaluated systems. Application-state propagation lets model operations and workflow steps share application state without supporting code. LangGraph and CrewAI expose workflow structure and state~\citep{langgraph,crewai}, while DSPy and NOOA provide declarative model interfaces~\citep{khattab2024dspy,nooa}. Other frameworks provide model--tool loops, typed interfaces, and multi-agent coordination~\citep{langchain,pydantic_ai,wu2024autogen,gao2024agentscope}.

Language model programming systems combine prompts with control flow and output constraints (LMQL)~\citep{beurerkellner2023lmql}, Python and asynchronous execution (APPL)~\citep{dong2025appl}, or generation and parallelism primitives (SGLang)~\citep{zheng2024sglang}. \sysname{} builds on MTP's typed model delegation~\citep{dantanarayana2025mtp} and OSP's graph traversal~\citep{mars2025osp}. This work defines common execution semantics for model operations, workflow structure and application state, and model-guided routing. The same workflow structure identifies the model's available choices and the steps the runtime can execute, exposing their relationships to compiler analysis. SIGIL compiles natural-language skills into Jac harnesses with typed model calls~\citep{dantanarayana2026sigil}. GraphMend extends Jac's compiler to remove PyTorch graph breaks~\citep{kashmira2025graphmend}.

Prior work optimizes prompts, code, and workflows from evaluation feedback~\citep{khattab2024dspy,agrawal2026gepa,he2025cognify,cheng2024trace,zhang2025aflow} and execution through parallelism, scheduling, and caching~\citep{kim2024llmcompiler,tan2025ayo,lin2024parrot,luo2026agentix,pan2025kvflow,ma2026mortis}. Speculative Actions validates predicted actions before commit~\citep{ye2026speculativeactions}; our pass identifies eligible requests and checks reuse at runtime. Service Weaver generates communication from component interfaces~\citep{ghemawat2023moderncloud}; our partitioner derives them from existing callables and configured boundaries.

%% file: sections/conclusion.tex
\section{Conclusion}
\label{sec:concl}

We introduce \textbf{Agent-Native Programming}, realized in \sysname{}, which unifies model operations, workflow structure and application state, and model-guided routing to simplify agent development and support automatic optimization. Across five benchmarks, \sysname{} reduces AI-integration code and coding-agent development effort, with competitive observed task quality. Case studies of speculative execution and service partitioning demonstrate how this unified execution model enables automatic optimizations through configuration alone, under stated conditions for preserving workflow behavior.

%% file: sections/statements.tex
\section*{AI Use}

We used Claude Code (Claude Opus 5.5) and Codex (GPT-6.1) to build, test, and revise applications in coding-agent experiments and configure or implement case-study optimizations. Benchmark applications used GLM-5.3, Claude Sonnet 5, and Gemma 4 31B; GPT-4.1-mini judged outputs for four benchmarks, and Jev provided auxiliary predictions for speculative execution.

\section*{Data-Availability Statement}

We will publicly release the \sysname{} implementation, benchmark applications, and evaluation code and data accompanying this work.